\documentclass[a4paper,11pt]{article}
\pdfoutput=1 
\usepackage{jheppub} 
\usepackage{bbm,bm,graphicx,mathtools,color,hyperref,slashed}
\usepackage{amssymb,amsmath}
\usepackage[T1]{fontenc}

\usepackage[all]{xy}

\graphicspath{{./Figures/}}

\newcommand{\diff}{\mathrm{d}}
\newcommand{\p}{\partial}

\newcommand{\be}{\begin{equation}}      
\newcommand{\ee}{\end{equation}}      
\newcommand{\bea}{\begin{eqnarray}}      
\newcommand{\eea}{\end{eqnarray}}

\newcommand{\tr}{\mathrm{tr}}

\newcommand{\im}{\mathrm{i}}

\newcommand{\calA}{\mathcal{A}}

\newcommand{\calZ}{\mathcal{Z}}
\newcommand{\rmc}{\mathrm{c}}
\newcommand{\rme}{\mathrm{e}}
\newcommand{\rmf}{\mathrm{f}}

\newcommand{\rmL}{\mathrm{L}}
\newcommand{\rmR}{\mathrm{R}}
\newcommand{\rmV}{\mathrm{V}}

\title{High-temperature domain walls of QCD with imaginary chemical potentials}

\author[1]{Hiromichi Nishimura,}

\emailAdd{hnishimura@bnl.gov}

\author[1,2]{Yuya Tanizaki}

\emailAdd{ytaniza@ncsu.edu}

\affiliation[1]{RIKEN BNL Research Center, Brookhaven National Laboratory, Upton, NY 11973 USA}

\affiliation[2]{Department of Physics, North Carolina State University, Raleigh, NC 27607, USA}

\abstract{
We study QCD with massless quarks on $\mathbb{R}^3\times S^1$ under symmetry-twisted boundary conditions with small compactification radius, i.e. at high temperatures. 
Under suitable boundary conditions, the theory acquires a part of the center symmetry and it is spontaneously broken at high temperatures. 
We show that these vacua at high temperatures can be regarded as different symmetry-protected topological orders, and the domain walls between them support nontrivial massless gauge theories as a consequence of anomaly-inflow mechanism. 
At sufficiently high temperatures, we can perform the semiclassical analysis to obtain the domain-wall theory, and $2$d $U(N_\mathrm{c}-1)$ gauge theories with massless fermions match the 't~Hooft anomaly. 
We perform these analysis for the high-temperature domain wall of $\mathbb{Z}_{N_\mathrm{c}}$-QCD and also of  Roberge-Weiss phase transitions. 
}

\begin{document}
\maketitle
\section{Introduction}\label{sec:introduction}

Confinement is one of the most important properties of non-Abelian gauge theory, and it still acquires a lot of interest to uncover its property. 
Although we cannot characterize confinement or deconfinement by using local order parameter, they are characterized by infrared behaviors of the loop operator, called Wilson loops, for certain $SU(N_\rmc)$ gauge theories~\cite{Wilson:1974sk, tHooft:1979rtg, tHooft:1981bkw}.  
In this sense, $SU(N_\rmc)$ Yang-Mills theory has the center symmetry $\mathbb{Z}_{N_\rmc}$, which is recently called $\mathbb{Z}_{N_\rmc}$ one-form symmetry~\cite{Gaiotto:2014kfa}. 
Spontaneous breakdown of higher-form symmetries imply the appearance of topological gauge theories in the infrared behaviors, and thus confinement and Higgs phases are separated as topological orders~\cite{PhysRevLett.96.110405, Kitaev:2005dm, Chen:2010gda}. 
In our real world, the strong-interaction sector of the Standard Model is described by quantum chromodynamics (QCD), which is the $SU(N_\rmc)$ gauge theory coupled to the Dirac fermions in the fundamental representation. 
In this case, since the color flux between two test quarks can break up by pair productions of dynamical quarks, we do not have a clear separation between confinement and Higgs phases as quantum phases of matters. In other words, the center symmetry is explicitly broken by the existence of dynamical quarks. 
Instead, QCD acquires chiral symmetry when quark masses are quite small, and chiral symmetry is spontaneously broken so as to generate mass scale~\cite{Nambu:1961tp, Nambu:1961fr}. 

These interesting behaviors, confinement and chiral symmetry breaking,  are the consequence of strong infrared dynamics, and it is usually very difficult to extract such information starting from QCD. 
An important direction is to perform the first-principle numerical computations of these systems, and the most established one is the Monte Carlo simulation of lattice gauge theories~\cite{Wilson:1974sk, Creutz:1980zw}. 
Another important direction is to discuss mathematically rigorous nature of QFTs. For this purpose, we have to find a quantity that is easily computable but is not affected by renormalization. 
Historically, this turned out to be true for 't~Hooft anomaly of QCD with massless quarks~\cite{tHooft:1979rat, Frishman:1980dq, Coleman:1982yg}, and we can conclude the existence of massless bound states when color degrees of freedom cannot be seen in the infrared. 
Thanks to the development of symmetry-protected topological phases~\cite{Vishwanath:2012tq, Wen:2013oza, Cho:2014jfa, Kapustin:2014lwa, Kapustin:2014zva, Wang:2014pma}, people understand that the applicability of 't~Hooft anomaly matching condition is much broader, and anomaly matching conditions are providing new insights on strongly-coupled QFTs~\cite{Witten:2015aba,Seiberg:2016rsg,Witten:2016cio, Tachikawa:2016cha, Tachikawa:2016nmo, Gaiotto:2017yup,Wang:2017txt, Tanizaki:2017bam, Komargodski:2017dmc, Komargodski:2017smk, Cho:2017fgz,Shimizu:2017asf, Wang:2017loc,Kikuchi:2017pcp, Gaiotto:2017tne, Gomis:2017ixy, Tanizaki:2017qhf, Tanizaki:2017mtm, Cherman:2017dwt, Yamazaki:2017dra, Guo:2017xex, Dunne:2018hog, Sulejmanpasic:2018upi, Cordova:2018cvg, Aitken:2018kky, Kobayashi:2018yuk, Tanizaki:2018xto, Cordova:2018acb, Anber:2018tcj, Anber:2018jdf, Tanizaki:2018wtg, Ohmori:2018qza, Hongo:2018rpy, Armoni:2018bga, Yonekura:2019vyz}. 

Using this development of knowledge, we study properties of confinement/deconfinement for QCD with massless quarks. 
Although the center symmetry does not exist for QCD as four-dimensional quantum field theories, we can find its interesting remnant by considering the symmetry-twisted boundary condition on the compactified spacetime $M_4=M_3\times S^1\ni (\bm{x},\tau)$. 
This is first discussed by Roberge and Weiss~\cite{Roberge:1986mm}: They consider QCD and introduce the non-thermal boundary condition for the quark field $\psi$ as 
\be
\psi(\bm{x},\tau+\beta)=\rme^{\im \phi}\psi(\bm{x},\tau). 
\ee 
Because of the gauge invariance, the partition function has the periodicity $2\pi/N_\rmc$ as a function of $\phi$, $\calZ(\phi+2\pi/N_\rmc)=\calZ(\phi)$, instead of the naive periodicity $2\pi$.  
This periodicity $2\pi/N_\rmc$ can be easily understood when quarks are confined inside hadrons. 
However, this periodicity is nontrivial when quarks are deconfined, and we have to introduce $N_\rmc$ branch structure of the free energy to make consistency. As a consequence, there exist the first-order phase transitions between $N_\rmc$ quasi-vacua at high temperatures, which are called Roberge-Weiss (RW) phase transitions. 
There are many studies on the property of RW phase transitions~\cite{Alford:1998sd, deForcrand:2002hgr,deForcrand:2003vyj,  DElia:2002tig, DElia:2004ani, Nagata:2011yf,  Nagata:2014fra, Bonati:2014kpa, Bonati:2016pwz, Bonati:2018fvg}.

Since the RW point has the first-order phase transition, we can consider the domain wall connecting those two pure states, and we will call it the high-temperature domain wall. 
High-temperature domain walls are recently studied for QCD with adjoint fermions~\cite{Anber:2018jdf, Anber:2018xek}, and the $2$d gauge theories coupled to chiral fermions appear on the domain wall. 
In adjoint QCD, the existence of chiral fermions on the domain wall is protected by mixed 't~Hooft anomaly between the center symmetry and the discrete chiral symmetry. 
In this paper, we will also find that the high-temperature domain wall at RW point supports $2$d gauge theories with massless fermions despite the fact that there is no one-form symmetry. 
We construct those gauge theories explicitly at sufficiently high temperatures, and compute the 't~Hooft anomaly for those theories on the domain wall. 

Our computation of the 't~Hooft anomaly of high-temperature domain walls suggest that the RW phase transition can be regarded as the phase transition between different symmetry-protected topological (SPT) orders. 
This new insight is consistent with the recent result in Refs.~\cite{Shimizu:2017asf, Yonekura:2019vyz}: the RW point has the mixed anomaly between ``RW parity'' symmetry and the chiral symmetry. 
One can also regard that our computation gives an explicit explanation about how the subtle parity anomaly found in Refs.~\cite{Shimizu:2017asf, Yonekura:2019vyz} is realized in the high-temperature QCD with imaginary chemical potential. 

In this paper, we also discuss the massless $\mathbb{Z}_N$-QCD~\cite{Kouno:2012zz, Sakai:2012ika, Kouno:2013zr, Kouno:2013mma, Poppitz:2013zqa, Iritani:2015ara, Kouno:2015sja, Hirakida:2016rqd, Hirakida:2017bye, Cherman:2017tey, Tanizaki:2017qhf, Tanizaki:2017mtm}, which is $SU(N)$ gauge theory with $N$-flavor massless fundamental quarks with symmetry twisted boundary condition, 
\be
\psi_{f}(\bm{x},\tau+\beta)=\rme^{2\pi\im f/N+\im \phi}\psi_f(\bm{x},\tau). 
\ee
This boundary condition is a special one since we have $\mathbb{Z}_N$ center-related symmetry as a three-dimensional QFT~\cite{Cherman:2017tey, Tanizaki:2017qhf, Tanizaki:2017mtm}, and we can discuss the domain wall connecting those vacua. 
Those domain walls are again given by $2$d $U(N-1)$ gauge theory coupled to massless fermions, but they produce the different chiral anomaly from that of RW domain wall. 
We will again see that it satisfies the anomaly-inflow mechanism from the bulk SPT phases by the help of the result in Refs.~\cite{Tanizaki:2017qhf, Tanizaki:2017mtm}. 

The paper is organized as follows. 
In Sec.~\ref{sec:domain_wall}, we give a brief description about the high-temperature domain wall in pure $SU(N_\rmc)$ Yang-Mills theory. 
In Sec.~\ref{sec:RW}, we discuss the massless gauge theories on high-temperature domain walls at the RW phase transition, and compute its 't~Hooft anomaly. 
In Sec.~\ref{sec:ZN_QCD}, we study the high-temperature domain walls of $\mathbb{Z}_N$-QCD. 
We summarize the result in Sec.~\ref{sec:conclusions}. 
In Appendix~\ref{sec:domain_wall_chirality}, we set our convention about the chirality of massless fermions on the domain wall. 
In Appendix~\ref{sec:ansatz_domain_wall}, we give justification of our ansatz about high-temperature domain wall used in this paper.

\section{Domain wall of $SU(N_\rmc)$ gauge theory in high-temperature phase}\label{sec:domain_wall}

Here, we consider the pure $SU(N_\rmc)$ Yang-Mills theory, for simplicity, and assume sufficiently high temperatures compared with the strong scale $\Lambda$, $T\gg \Lambda$. At high-temperature, $\mathbb{Z}_{N_\rmc}$ center symmetry is spontaneously broken and there are $N_\rmc$ discrete vacua, described by the expectations values of the Polyakov loop, 
\be
\Phi(\bm{x})=\mathcal{P}\exp\int_{0}^{\beta} a_4(\bm{x},\tau) \diff \tau. 
\ee
Let us take the Polyakov gauge~\cite{Reinhardt:1997rm}, and then the effective action becomes
\be
S_{1\mathrm{-loop}}={\beta\over g^2}\int_{\mathbb{R}^3}\diff^3\bm{x} \left({1\over 2}\tr[F_{IJ}^2]+\tr[(D_I a_4)^2]+g^2 V(a_4)\right), 
\ee
with the one-loop effective action~\cite{Gross:1980br, Weiss:1980rj}
\be
V(a_4)=-{2\over \pi^2 \beta^4}\sum_{n\ge 1}{1\over n^4}\left(\Bigl|\tr_\rmc(\Phi^n)\Bigr|^2-N_\rmc^2\right). 
\label{V_1loop}
\ee
We here give an offset to $V(a_4)$ so that $V(a_4)\ge 0$ and $V(0)=0$. 
This $3$-dimensional theory is again a strongly-coupled non-Abelian gauge theory, and we cannot solve it analytically. 
Throughout this paper, we assume the standard lore saying that $3$-dimensional gluons get mass gap and the spatial Wilson loop shows area law, and the $\mathbb{Z}_{N_\rmc}$ zero-form symmetry $\Phi\mapsto \rme^{2\pi\im/N_\rmc}\Phi$ is spontaneously broken as suggested by the one-loop effective action.  

There are $N_\rmc$ vacua characterized by the Polyakov-loop expectation values, 
\be
{1\over N_\rmc}\langle \tr(\Phi)\rangle =\rme^{2\pi\im k/N_\rmc}, 
\ee
with $k=0,1,\ldots, N_\rmc-1$. 
We can consider a domain wall connecting these pure states, and call it a high-temperature domain wall. 
We put the following ansatz of the high-temperature domain wall~\cite{Bhattacharya:1992qb}, 
\be
\Phi(\bm{x})=\exp\left(\im \rho(x_3) T_{N_\rmc-1} \right), 
\ee
where $T_{N_\rmc-1}={2\pi\over N_\rmc}\mathrm{diag}(1,\ldots,1,1-N_\rmc)$ is the last Cartan element of $\mathfrak{su}(N_\rmc)$, and $\rho(x_3)$ should be determined by the classical equation of motion with the boundary condition $\rho(x_3=-\infty)=0$ and $\rho(x_3=\infty)=1$. This domain wall connects the vacua $k=0$ and $k=1$. 
We can discuss the BPS bound~\cite{Bogomolny:1975de, Prasad:1975kr} in terms of $\rho(x_3)$ within this ansatz. 
Justification of this ansatz will be discussed in Appendix~\ref{sec:ansatz_domain_wall}. 

With this ansatz, the adjoint Higgsing 
\be
SU(N_\rmc)\to U(N_\rmc-1)=[SU(N_\rmc-1)\times U(1)]/\mathbb{Z}_{N_\rmc-1}
\ee
occurs near the domain wall, $\rho\simeq 1/2$. The division by $\mathbb{Z}_{N_\rmc-1}$ can be understood as follows: The embedding $SU(N_\rmc-1)\times U(1)\hookrightarrow SU(N_\rmc)$ is given by
\be
(U_{N_\rmc-1},\mathrm{e}^{\im \phi})\mapsto 
\begin{pmatrix}
\rme^{\im \phi} U_{N_\rmc-1}&0\\
0&\rme^{-\im (N_\rmc-1)\phi}
\end{pmatrix}. 
\ee
The kernel of this embedding is given by $\mathbb{Z}_{N_\rmc-1}$, whose generator is given by the element $(\rme^{2\pi\im/(N_\rmc-1)}\bm{1}_{N_\rmc-1}, \rme^{-2\pi\im/(N_\rmc-1)})$, and thus the image of the map inside $SU(N_\rmc)$ is isomorphic to $[SU(N_\rmc-1)\times U(1)]/\mathbb{Z}_{N_\rmc-1}$. 
The defining representation $\bm{N}_\rmc$ breaks up into $(\bm{N_\rmc-1})_{+1}\oplus (\bm{1})_{-(N_\rmc-1)}$, where $(\bm{R})_{Q}$ denotes the representation $\bm{R}$ of $SU(N_\rmc-1)$ with the $U(1)$ charge $Q$. The possible value of $Q$ is constrained by the $(N_\rmc-1)$-ality of $\bm{R}$.

Although we obtain a nontrivial gauge theory on the domain wall, we do not expect that it causes any interesting low-energy physics like spontaneous symmetry breaking after taking into account the quantum fluctuation because even $2$d $U(1)$ pure gauge theory acquires the trivial mass gap. 
Only exception would be the case with topological term with $\theta=\pi$, as discussed in Refs.~\cite{Gaiotto:2017yup, Anber:2018xek}. 
This is a good lesson for us; even if we find the massless Lagrangian on the domain wall,  it is important to discuss whether that massless nature survives under quantum and thermal fluctuations (see also Refs.~\cite{Anber:2015kea, Sulejmanpasic:2016uwq}). 
This motivates us to compute the 't~Hooft anomaly of the domain wall theory~\cite{Komargodski:2017smk}. 

\section{Domain wall at the Roberge-Weiss phase transitions}
\label{sec:RW}

In this section, we consider QCD with quark imaginary chemical potential $\phi/\beta$, which is related to the $U(1)_\rmV$-twisted boundary condition on the quark field, $\psi(\tau+\beta)=\rme^{\im\phi}\psi(\tau)$. Roberge and Weiss showed that the QCD partition function has a fractional periodicity $\phi\sim \phi+2\pi/N_\rmc$ by the gauge invariance~\cite{Roberge:1986mm}. 
This Roberge-Weiss periodicity concludes that these is a $\mathbb{Z}_2$ center symmetry for quantized values of $\phi$~\cite{Shimizu:2017asf, Yonekura:2019vyz}, and we call it the RW parity. 
At high temperatures, this RW parity is spontaneously broken, and there are two degenerate vacua with the mass gap as the three-dimensional quantum field theory. 
We show that these distinct vacua can be regarded as the different SPT phases by studying the domain wall between them. The RW domain wall supports massless field theories by anomaly-inflow arguments.

\subsection{Imaginary chemical potential and Roberge-Weiss phase transition}

Let us start with a review of RW phase transition~\cite{Roberge:1986mm}. We consider the QCD partition function with $U(1)$ imaginary chemical potential, 
\be
Z(T,\phi)=\tr\left[\exp\left(-\beta (\widehat{H}+\im \mu_{\mathrm{I}} \widehat{Q})\right)\right], 
\ee
where $\widehat{H}$ is the QCD Hamiltonian with massless quarks, $\widehat{Q}$ is the quark number operator, $T=1/\beta$ is the temperature, $\mu_{\mathrm{I}}$ is the imaginary chemical potential, and we set $\phi=\beta \mu_{\mathrm{I}}+\pi$. 
In the path-integral expression, this is the $SU(N_\rmc)$ Yang-Mills theory coupled to $N_\rmf$-flavor massless Dirac fermions, 
\be
S={1\over g^2}\int \tr[F_\rmc\wedge \star F_\rmc]+\int \diff^4 x \sum_{f=1}^{N_\rmf}\overline{\psi}_f\gamma_{\mu}\left(\p_{\mu}+a_{\mu}\right)\psi_f, 
\ee
where $a$ is the $SU(N_\rmc)$ gauge field, $F_\rmc=\diff a +a\wedge a$ is the $SU(N_\rmc)$ field strength, and $\psi_f,\overline{\psi}_f$ are the four-dimensional Dirac fermions with the flavor label $f=1,\ldots,N_\rmf$. Our spacetime is $\mathbb{R}^3\times S^1$, and the boundary condition of the quark fields is twisted by $U(1)$ phase,
\be
\psi(\bm{x},\tau+\beta)=\mathrm{e}^{\im\phi}\psi(\bm{x},\tau). 
\label{eq:boundary_condition_RW}
\ee
Naively, we expect the periodicity $\phi\sim \phi+2\pi$ from this expression. Since the gauge-invariant operators have the charge in $N_\rmc\mathbb{Z}$, however, the partition function has a shorter periodicity $\phi\sim \phi+2\pi/N_\rmc$. 
In other words, we can consider the transformation, $\Phi(x)\mapsto \rme^{2\pi \im/N_\rmc}\Phi(x)$, on the Polyakov loop associated with simultaneously change the boundary condition of quark fields as 
\be
\psi(\bm{x},\tau+\beta)=\mathrm{e}^{\im(\phi+2\pi /N_\rmc)}\psi(\bm{x},\tau), 
\ee
and then the value of the partition function does not change. 
This is called the Roberge-Weiss (RW) periodicity~\cite{Roberge:1986mm}. 
Here, it is important to notice that the above transformation is \textit{not} symmetry because the boundary condition of matter fields are different. RW periodicity just says that the theory with two different boundary conditions have the same free energy.

The continuous symmetry of massless QCD is given by~\cite{Shimizu:2017asf,  Tanizaki:2017qhf, Tanizaki:2017mtm, Tanizaki:2018wtg, Yonekura:2019vyz}
\be
G={SU(N_\rmf)_\rmL\times SU(N_\rmf)_\rmR\times U(1)_\rmV\over \mathbb{Z}_{N_\rmc}\times \mathbb{Z}_{N_\rmf}}. 
\ee
For generic values of $\phi$, the charge conjugation $\mathsf{C}:\psi\mapsto \mathcal{C} \overline{\psi}^t$ with $\mathcal{C}=\im \gamma_2\gamma_4$ is explicitly broken by the boundary condition, because (\ref{eq:boundary_condition_RW}) is changed as 
\be
\psi(\bm{x},\tau+\beta)=\mathrm{e}^{-\im\phi}\psi(\bm{x},\tau). 
\ee
If $\phi$ is quantized to $\pi/N_\rmc$, we can construct the charge-conjugation symmetry by simultaneously inserting the appropriate color 't~Hooft magnetic flux~\cite{Shimizu:2017asf, Yonekura:2019vyz}\footnote{In Ref.~\cite{Yonekura:2019vyz}, the author considered the $\mathsf{PT}$ transformation $\mathsf{R}:(x_3,\tau)\mapsto (-x_3,-\tau)$ instead of $\mathsf{C}$. These two are equivalent because of $\mathsf{CPT}$ theorem. }: When $\phi=- (\pi/N_\rmc)k$, we define the following $\mathbb{Z}_2$ transformation, 
\be
\Phi(\bm{x})\mapsto \rme^{2\pi\im k/N_\rmc}{\Phi}(\bm{x})^\dagger,\; 
\psi(\bm{x},\tau)\mapsto \rme^{(2\pi\im/N_\rmc )k}\mathcal{C}\overline{\psi}(\bm{x},\tau)^t. 
\ee
We call this $\mathbb{Z}_2$ transformation as the RW parity. After this transformation, the boundary condition is changed as 
\be
\psi(\bm{x},\tau+\beta)=\mathrm{e}^{-2\pi\im k/N_\rmc-\im\phi}\psi(\bm{x},\tau), 
\ee
which is the same with (\ref{eq:boundary_condition_RW}) when $\phi=-\pi k/N_\rmc $. Therefore, the RW parity is the symmetry of the theory, and thus the symmetry group is enhanced at $\phi=-\pi k/N_\rmc $ as 
\be
G\rtimes (\mathbb{Z}_2)_{\mathrm{RW}}. 
\ee

Now, let us discuss the high-temperature phase of massless QCD with imaginary chemical potential. 
High-$T$ behavior is controlled by the one-loop potential, and they are given by
\bea
V_{\mathrm{gluon}}&=&-{2\over \pi^2 \beta^4}\sum_{n\ge 1}{1\over n^4}\left(\Bigl|\tr_\rmc(\Phi^n)\Bigr|^2-1\right),\\
V_{\mathrm{quark}}&=&{2 N_\rmf\over \pi^2 \beta^4}\sum_{n\ge 1}{1\over n^4}\left(\rme^{\im n\phi}\tr_\rmc(\Phi^n)+\rme^{-\im n \phi}\tr_\rmc( (\Phi^\dagger)^n)\right). 
\eea
We immediately see that, at $\phi=-\pi-{2\pi\over N_\rmc}k$, the classical vacuum $\Phi_k=\rme^{2\pi\im k/N_\rmc}\bm{1}$ is chosen. 
At the intermediate point, $\phi=-\pi-{\pi\over N_\rmc}$, the vacua connected to $k=0$ and $k=1$ has the same lowest energy, and we have the first-order phase transition between those two states: this is called the RW phase transition. Since these pure states $k=0$ and $k=1$ are related by $(\mathbb{Z}_2)_{\mathrm{RW}}$, the RW phase transition is the consequence of spontaneous breakdown of RW parity symmetry. 

\subsection{Semiclassical analysis of domain wall theory at $T\gg \Lambda$}

At really high temperatures, $T\gg \Lambda$, we can do a certain semi-classical calculations due to asymptotic freedom, and obtain the three-dimensional effective theory as we have explained. Strictly speaking, we cannot still solve the problem even in that regime, since such a dimensionally-reduced theory is again typically strongly coupled and cannot be solved. 
Therefore, let us adopt a standard lore that the three-dimensional Yang-Mills theory (without Chern-Simons terms) is trivially gapped with the mass scale $\sim g^2 T$, and we will derive the nontrivial domain-wall theory under this assumption.

Since the RW parity is spontaneously broken at the RW phase transition, we can consider the high-temperature domain wall connecting two pure states. 
In the following, we take the RW point,
\be
\phi=-\pi-{\pi\over N_\rmc},
\ee
and consider the domain wall connecting $k=0$ and $k=1$:
\be
\Phi(x_3)=\exp\left(\im \rho(x_3)T_{N_\rmc-1}\right), 
\ee
with $\rho(-\infty)=0$ and $\rho(\infty)=1$.

The $4$d kinetic term of the $f$-th flavor fermion is given by 
\be
\overline{\psi}_f \gamma_I D_I \psi_f+\overline{\psi}_f\gamma_4\left(\p_4+{\im\over \beta} \rho(x_3)T_{N_\rmc-1}\right)\psi_f, 
\ee
and the imaginary-time direction gives the real mass term for $3$d fermions with the mass
\be
m_{n}(x_3)=2\pi n+\rho(x_3)T_{N_\rmc-1}-\pi-{\pi\over N}. 
\ee
Under the Higgsing $SU(N_\rmc)\to [SU(N_\rmc-1)\times U(1)]/\mathbb{Z}_{N_\rmc-1}$, the fermion is decomposed into $({\bm{N_\rmc-1}})_1$ and $(\bm{1})_{-(N_\rmc-1)}$. 
The mass function for $({\bm{N_\rmc-1}})_1$ is 
\be
m^{(\bm{N_\rmc-1})}_n(x_3)= 2\pi n+{2\pi\over N}\rho(x_3)-\pi-{\pi\over N}\not=0. 
\ee
Thus, $({\bm{N_\rmc-1}})_1$ completely decouples in the low-energy limit. The mass function for $(\bm{1})_{-(N_\rmc-1)}$ is 
\be
m^{(\bm{1})}_n(x_3)=2\pi n-{2\pi\over N}(N-1)\rho(x_3)-\pi-{\pi\over N}. 
\ee
For $n=1$, this takes zero at $x_3=0$, and others cannot be zero. As a consequence, we get $N_\rmf$-flavor $2$d Dirac fermions with the gauge charge $(\bm{1})_{-(N_\rmc-1)}$ living on the domain wall (see Appendix~\ref{sec:domain_wall_chirality} for details). 

The quark kinetic term on the domain wall is given by 
\be
\sum_{f=1}^{N_\rmf}\overline{\psi}_f^{(\bm{1})} \bigl(\sigma_I [\p_I -\tr(a'_I)] \bigr) \psi^{(\bm{1})}_f, 
\ee
where $a'$ is the $U(N_\rmc-1)$ gauge field and $\psi_f^{(\bm{1})}$ is the $2$d massless Dirac fermion in the representation $(\bm{1})_{-(N_\rmc-1)}$, which comes out of the normalizable zero mode of $4$d Dirac fermion $\psi_f$. 
This low-energy theory has the chiral symmetry, 
\be
{SU(N_\rmf)_\rmL\times SU(N_\rmf)_\rmR\over \mathbb{Z}_{N_\rmf}}\subset G. 
\ee
The $U(1)$ baryon number symmetry in $G$ cannot be seen within this low-energy Lagrangian since the fermions in the $(\bm{N_\rmc-1})_{1}$ representation is completely neglected because of their nonzero thermal mass. 
This two-dimensional field theory has the chiral anomaly characterized by three-dimensional level-$1$ Chern-Simons action, 
\be
\mathrm{CS}_3[L]-\mathrm{CS}_3[R]={1\over 4\pi}\tr\left(L\diff L+{2\over 3}L^3\right)-{1\over 4\pi}\tr\left(R\diff R+{2\over 3}R^3\right),  
\label{eq:anomaly_RW_domain_wall}
\ee
where $L$ and $R$ are background gauge fields for $SU(N_\rmf)_\rmL$ and $SU(N_\rmf)_\rmR$, respectively. 

Here, we demonstrate that the high-temperature domain wall at the RW phase transition supports $U(N_\rmc-1)$ gauge theory coupled to $N_\rmf$ massless $2$d Dirac fermions by semiclassical analysis at sufficiently high temperatures. 
We expect that the validity of that effective theory is limited to sufficiently high temperatures because the theory is expected to become strongly coupled near the chiral restoration temperatures. 
The anomaly, however, is a topological obstruction of gauging the global symmetry and it cannot be changed under local deformation of the Lagrangian so long as the symmetry is respected. 
This suggests that, so long as the RW parity is spontaneously broken, the high-$T$ domain wall should supports the $2$d massless field theory with an 't~Hooft anomaly characterized by (\ref{eq:anomaly_RW_domain_wall}). 
Since the same anomaly is carried by the level-$1$ $SU(N_\rmf)$ Wess-Zumino-Witten model, it seems to be natural to expect that the long-range behavior on the high-$T$ domain wall is given by that conformal field theory. 

\subsection{Anomaly inflow from $3$d bulk}\label{sec:anomaly_inflow_3d_RW}

In this section, we confirm more explicitly that the domain-wall theory should have an anomaly (\ref{eq:anomaly_RW_domain_wall}) by using the anomaly-inflow mechanism~\cite{Callan:1984sa}. 
Recent understanding of the 't~Hooft anomaly says that the system with 't~Hooft anomaly should be realized as a boundary of symmetry-protected topological (SPT) orders if anomalous symmetry is weakly gauged~\cite{Vishwanath:2012tq, Wen:2013oza, Cho:2014jfa, Kapustin:2014lwa}. 
Then, the anomaly of the boundary theory is canceled by anomaly inflow from the bulk SPT order, and the combined system has no anomaly. 
In the case of the RW phase transition, since $3$-dimensional theory has no chiral anomaly, the bulk gapped states separated by the high-temperature domain wall can be regarded as different SPT phases protected by the chiral symmetry. 

The above anomaly-inflow discussion has a nice consistency with the recent discussion in Refs.~\cite{Shimizu:2017asf, Yonekura:2019vyz}. 
These papers show that the RW point has a mixed 't~Hooft anomaly between the RW parity and the chiral symmetry: Let us consider the QCD partition function $\calZ_{\mathrm{RW}}[L,R]$ on $\mathbb{R}^3\times S^1$ with the background $3$-dimensional $SU(N_\rmf)_{\rmL,\rmR}$ gauge fields $L,R$. 
With these backgrounds, the RW parity at $\phi=-\pi-\pi/N_\rmc$ is anomalously broken, 
\be
(\mathbb{Z}_2)_{\mathrm{RW}}: \calZ_{\mathrm{RW}}[L,R]\mapsto \calZ_{\mathrm{RW}}[L,R]\exp\left(\mathrm{CS}_3[L]-\mathrm{CS}_3[R]\right). 
\ee
This relation shows that, when RW parity is broken, the partition functions of those two pure states are different by the Chern-Simons action, $\exp\left(\mathrm{CS}_3[L]-\mathrm{CS}_3[R]\right)$. 
Since we are assuming that the $3$d bulk is gapped, this justifies that these two pure states are different as SPT orders protected by the chiral symmetry. 

\section{Domain wall of massless $\mathbb{Z}_N$-QCD at high temperatures}
\label{sec:ZN_QCD}

In this section, we consider massless QCD with $N_\rmc=N_\rmf=N$, and we take the flavor-dependent boundary condition. This theory has the color-flavor locked center symmetry $\mathbb{Z}_N$, and it is called $\mathbb{Z}_N$-QCD~\cite{Kouno:2012zz, Sakai:2012ika, Kouno:2013zr, Kouno:2013mma, Poppitz:2013zqa, Iritani:2015ara, Kouno:2015sja, Hirakida:2016rqd, Hirakida:2017bye, Cherman:2017tey, Tanizaki:2017qhf, Tanizaki:2017mtm}. 
At high temperatures, this center symmetry is spontaneously broken, and there are $N$ distinct vacua with the mass gap as the three-dimensional quantum field theory. 
As in the case of the RW phase transition, we show that these distinct vacua can be regarded as the different SPT phases. As a consequence, the domain wall connecting them support massless field theories by anomaly-inflow arguments.

\subsection{Massless $\mathbb{Z}_N$-QCD and center symmetry}

$\mathbb{Z}_N$-QCD is $SU(N)$ gauge theory with degenerate $N$-flavor fundamental quarks with both the flavor-twisted and $U(1)_\rmV$-twisted boundary conditions:
\be
\psi_{f}(\bm{x},\tau+\beta)=\exp\left(\im{2\pi\over N}f+\im\phi\right)\psi_f(\bm{x},\tau)
\ee
where $f= 1, \dots, N$.
As we show below, this theory has the color-flavor locked $\mathbb{Z}_N$ center symmetry, so it is called $\mathbb{Z}_N$-QCD~\cite{Kouno:2012zz, Sakai:2012ika, Kouno:2013zr, Kouno:2013mma, Poppitz:2013zqa, Iritani:2015ara, Kouno:2015sja, Hirakida:2016rqd, Hirakida:2017bye, Cherman:2017tey, Tanizaki:2017qhf, Tanizaki:2017mtm}.  We take the fermion mass to be zero in this paper. 

Let us give a detailed comment on the symmetry of massless $\mathbb{Z}_N$-QCD. We start with the internal symmetry of massless QCD for generic numbers of color $N_\rmc$ and flavor $N_\rmf$:
\be
G={SU(N_\rmf)_\rmL\times SU(N_\rmf)_\rmR\times U(1)_\rmV\over \mathbb{Z}_{N_\rmc}\times \mathbb{Z}_{N_\rmf}} . 
\ee
Representing the Dirac fields in the chiral basis, $\psi=(\psi_\rmR,\psi_\rmL)$, $SU(N_\rmf)_\rmL\times SU(N_\rmf)_\rmR\times U(1)_\rmV$ acts on the quark field $\psi$ as 
\be
(g_{\rmL},g_{\rmR},\mathrm{e}^{\im\alpha}): \psi\mapsto \mathrm{e}^{\im\alpha}(g_\rmR \psi_\rmR, g_{\rmL}\psi_\rmL). 
\ee
Since $(g_\rmL,g_\rmR,\mathrm{e}^{\im\alpha})$ and $(g_{\rmL}\rme^{2\pi\im/N_\rmf},g_{\rmR}\rme^{2\pi\im/N_\rmf},\mathrm{e}^{\im\alpha-2\pi\im/N_\rmf})$ give the same mapping, the symmetry group must be divided by $\mathbb{Z}_{N_\rmf}$ to remove this redundancy. 
Furthermore, any gauge-invariant local operator has the quantized charge $N_\rmc$ under $U(1)_\rmV$,  and thus we also have to introduce the identification, $\alpha\sim \alpha+2\pi/N_\rmc$, and obtain the above symmetry group. 
We are interested in the subgroup of $G$:
\be
G^{\mathrm{sub}}={SU(N_\rmf)_\rmV\times U(1)_\rmV\over \mathbb{Z}_{N_\rmc}\times \mathbb{Z}_{N_\rmf}}\times (\mathbb{Z}_{N_\rmf})_\rmL. 
\ee
Let us now set $N_\rmc=N_\rmf=N$ and consider the symmetry of $\mathbb{Z}_N$-QCD. The vector-like flavor symmetry $SU(N)$ is broken down to $U(1)^{N-1}$ because of the flavor-dependent boundary conditions, so naively the symmetry group seems to become 
\be
{U(1)^N\over (\mathbb{Z}_N)_\rmc}\times (\mathbb{Z}_N)_\rmL, 
\ee
but there is an extra $\mathbb{Z}_N$ symmetry, called the color-flavor center symmetry~\cite{Cherman:2017tey,Tanizaki:2017qhf, Tanizaki:2017mtm}. 

The center symmetry of the pure Yang-Mills theory is the $\mathbb{Z}_N$ one-form symmetry acting on the Wilson lines. On $\mathbb{R}^3\times S^1$, it induces the $\mathbb{Z}_N$ zero-form symmetry acting on the Polyakov loop $\Phi(\bm{x})$ as 
\be
\Phi(\bm{x})\mapsto \mathrm{e}^{2\pi\im/N}\Phi(\bm{x}). 
\ee
This symmetry is explicitly broken in QCD, because it changes the boundary condition of the quark field as 
\be
\psi_{f}(\bm{x},\tau+\beta)=\rme^{2\pi\im/N}\exp\left(\im{2\pi\over N}f+\im\phi\right)\psi_f(\bm{x},\tau). 
\ee
For $\mathbb{Z}_N$-QCD, however, one can compensate this violation by performing the shift of the flavor label, $\psi_{f}\mapsto \psi_{f+1}$, which is a part of the vector-like $SU(N_\rmf)\times U(1)$ symmetry. Therefore, $\mathbb{Z}_N$-QCD has a symmetry generated by~\cite{Cherman:2017tey, Tanizaki:2017qhf, Tanizaki:2017mtm} 
\be
\Phi(\bm{x})\mapsto \rme^{2\pi\im/N}\Phi(\bm{x}),\; \psi_{f}\mapsto \psi_{f+1}, 
\ee
and we call it the center symmetry $(\mathbb{Z}_N)_{\mathrm{center}}$. The symmetry group of massless $\mathbb{Z}_N$-QCD is thus obtained:
\be
(\mathbb{Z}_{N})_{\mathrm{center}}\ltimes {U(1)^N\over \mathbb{Z}_N}\times (\mathbb{Z}_{N})_\rmL. 
\ee
The noncommutativity between $(\mathbb{Z}_{N})_{\mathrm{center}}$ and $U(1)^N$ originates from the fact that $U(1)^N$ is the maximal Abelian subgroup of $U(N)$ and the shift of the flavor label is given by the non-diagonal matrix of $U(N)$.

\subsection{Semiclassical analysis of the domain wall theory at $T\gg\Lambda$}\label{sec:DW}

Let us make the quark field periodic by mapping $\psi_f(\tau)\mapsto \exp\left({\im\over \beta} \left({2\pi\over N}f+\phi\right)\tau\right)\psi_f(\tau)$.  
Then, the Dirac operator becomes 
\be
\sum_{I=1}^{3}\overline{\psi}_f \gamma_I D_I \psi_f+\overline{\psi}_f\gamma_4\left(\p_4+a_4+\im{2\pi\over N\beta}f+{\im \phi\over \beta}\right)\psi_f. 
\ee
We take the Polyakov gauge, so that $a_4$ is diagonal and $\tau$-independent. We take a domain-wall solution, which connects two perturbative vacuum of the gluon one-loop potential, $\Phi=\bm{1}$ and $\Phi=\rme^{2\pi\im/N}\bm{1}$, as 
\be
\langle \Phi(x_3)\rangle = \langle \rme^{\im \beta a_4(x_3)}\rangle=\exp\left(\im \rho(x_3)T_{N-1}\right), 
\ee
where $T_{N-1}={ 2\pi \over N} \mathrm{diag}(1,\ldots,1,1-N)$.\footnote{Exact expression or exact location of the classical vacuum does not matter in the following argument. But we point out that the gluon potential is $O(N^2)$ while the quark potential is $O(N^{-2})$ under this twisted boundary condition~\cite{Cherman:2017tey}, and there are the factor $100$ difference already for $N=3$. So the use of the classical vacuum of gluon potential should be a good approximation.} 

Since the fluctuation of $a_4$ should be small at high temperatures, we can take
\be
\langle a_4(x_3)\rangle=\im {\rho(x_3) T_{N-1}\over \beta}. 
\ee
Therefore, the quark kinetic term becomes 
\bea
\sum_{i=1}^{2}\overline{\psi}_f \gamma_i D_i \psi_f
+\overline{\psi}_f \left[\gamma_3\p_3+{2\pi\im\over \beta}\gamma_4\left(n+{\rho(x_3) \over 2 \pi}T_{N-1}+{f\over N}+{\phi\over 2\pi}\right)\right]\psi_f. 
\eea
As we mentioned, we want to take a special $\phi$ so that the real mass is non-zero for any $n$ and $f=0,\ldots,N-1$ when $\rho(x_3)=0$. As such an example, let us take 
\be
\phi=-\pi/N, 
\ee
then the real mass on the bulk becomes 
\be
{2\pi\over \beta}\left(n+{2f-1\over 2N}\right)\not=0. 
\ee
This ensures that the quarks are classically massive as it acquires the real mass, $m\gtrsim {1\over N}\pi T$. 

The gauge group $SU(N)$ is Higgsed to $[SU(N-1)\times U(1)]/\mathbb{Z}_{N-1}$ near the domain wall.
The fundamental quark in the representation $\bm{N}$ is thus breaks into $(\bm{N-1})_{1}$ and $(\bm{1})_{-(N-1)}$.  
First $(N-1)$ color of fermions (i.e. $({\bm{N-1}})_1$) have the mass
\be
m_{n,f}^{(\bm{N-1})_1}(x_3)={2\pi\over N\beta}\left(N n + \rho(x_3)+f-{1\over 2}\right), 
\ee
with $f=0,1,\ldots, N-1$ and $0=\rho(-\infty)\le \rho(x_3)\le \rho(\infty)=1$. The mass function flip its sign only for $n=0$ and $f=0$, and others have the definite sign. Therefore, only the mode with $n=0$ and $f=0$ can be a candidate of the domain-wall fermions with the gauge representation $(\bm{N-1})_{1}$. 
Second the last color component of the fermion (i.e. $(\bm{1})_{-(N-1)}$) has the mass function 
\be
m_{n,f}^{(\bm{1})_{-(N-1)}}={2\pi\over N\beta}\left(N n-\rho(x_3)(N-1)+f-{1\over 2}\right). 
\ee
For $n\not=0$, this always has the definite mass, and so does for $f=0$. Thus the candidate of the domain wall fermions are $n=0$ and $f=1,\ldots,N-1$. 

Since the direction of the sign of those domain wall masses are flipped between $(\bm{N-1})_1$ and $(\bm{1})_{-(N-1)}$, the chirality between these two representations of $2$D fermions are opposite. 
For the convention of the chirality, one can see Appendix~\ref{sec:domain_wall_chirality}.  
Thus, the gauged quark kinetic term now becomes 
\bea
&&\overline{\psi}_0^{\bm{N-1}} \sigma_i(\p_i+a'_i+A_{0,i}-A_{\chi,i}P_\rmL)\psi_0^{\bm{N-1}}\nonumber\\
&&+{\sum_{f'=1}^{N-1}}\overline{\psi}_{f'}^{\bm{1}}\sigma_i(\p_i-\tr[a'_i]+A_{f',i}+A_{\chi,i}P_\rmL) \psi_{f'}^{\bm{1}},
\eea
where $a'$ is the $U(N-1)$ dynamical gauge field, $A_f$ are $U(1)$ background gauge fields for $f$-th flavor rotation, and $A_\chi$ is the $\mathbb{Z}_N$ background gauge field for the discrete chiral symmetry. 
We can find that the mixed anomaly between the gauge symmetry $U(N-1)$ and the discrete chiral symmetry $(\mathbb{Z}_N)_\rmL$ cancels among $N$ domain-wall fermions, and then we find the following SPT action for the 't~Hooft anomaly of the domain-wall theory,
\be
S_{\mathrm{SPT,DW}}=\sum_{f=0}^{N-1}{1\over 2\pi}\int A_\chi\wedge \diff A_f. 
\label{eq:anomaly_ZN_QCD_wall}
\ee
In order to find this result, it is convenient to use the Stora-Zumino chain. The starting point is the $4$-dimensional Abelian anomaly $\calA_4$; 
\bea
\calA_4&=&{2\pi\over 2! (2\pi)^2}\int \tr_{(N-1)}\left[(F'+\diff A_0-\diff A_\chi)^2-(F'+\diff A_0)^2\right]\nonumber\\
&&+\sum_{f'=1}^{N-1}{2\pi\over 2! (2\pi)^2}\int \left[(-\tr F'+\diff A_{f'}+\diff A_\chi)^2-(- \tr F'+\diff A_{f'})^2\right]\nonumber\\
&=&\sum_{f=0}^{N-1}{1\over 2\pi}\int \diff A_\chi\wedge \diff A_f-{1\over 2\pi}\int N \diff A_\chi\wedge \left(\tr F'+\diff A_0\right). 
\eea
Here, $F'=\diff a'+a'\wedge a'$ is the $U(N-1)$ field strength, and the second term of the last line vanishes modulo $2\pi$. We obtain the $3$-dimensional topological action $S_{\mathrm{SPT,DW}}$ as a boundary theory of $\calA_4$. 

Here, we elucidated that the domain wall connecting different vacua related by $(\mathbb{Z}_N)_{\mathrm{center}}$ supports the $(1+1)$-dimensional gauge theory with massless Dirac fermions. 
The computation is done in the semiclassical regime, $T\gg \Lambda$, but we can argue its persistence because of the topological nature of anomaly. 
In other words, the gapped vacua related by $(\mathbb{Z}_N)_{\mathrm{center}}$ are different as symmetry-protected topological order, and the difference is characterized by the $\mathbb{Z}_N$ topological action (\ref{eq:anomaly_ZN_QCD_wall}). 
Let us emphasize that this facts survive even at $T\gtrsim \Lambda$ so long as the system is in the deconfined phase.

Moreover, we can also prove this statement from the anomaly-inflow mechanism as we have done for the Roberge-Weiss high-temperature domain wall in Sec.~\ref{sec:anomaly_inflow_3d_RW}. 
In Refs.~\cite{Tanizaki:2017qhf, Tanizaki:2017mtm}, it is found that $\mathbb{Z}_N$-QCD has the mixed 't~Hooft anomaly among $(\mathbb{Z}_N)_{\mathrm{center}}$, $U(1)^N/\mathbb{Z}_N$, and $(\mathbb{Z}_N)_\rmL$ symmetries. 
In our context, it is useful to summarize this result as the partition function $\calZ_{\mathbb{Z}_N}$ of $\mathbb{Z}_N$-QCD breaks $(\mathbb{Z}_N)_{\mathrm{center}}$ symmetry anomalously under the existence of background gauge fields:
\be
(\mathbb{Z}_N)_{\mathrm{center}}:\calZ_{\mathbb{Z}_N}[A_f,A_\chi]\mapsto \calZ_{\mathbb{Z}_N}[A_f,A_\chi]\exp\left(\im S_{\mathrm{SPT,DW}}[A_f,A_\chi]\right).
\ee
This says that the $3$d high-temperature states related by the broken center symmetry can be regarded as the different symmetry protected topological states, and the domain wall between them should cancel the anomaly inflow from the bulk. 
This argument does not use any concrete information of the construction of domain-wall theories, and thus it shows the robustness of the existence of nontrivial ground states under the effect of quantum and thermal fluctuations. 

\section{Summary}\label{sec:conclusions}

We study the domain-wall localized theories at the high-temperature phase of QCD with massless fundamental quarks under symmetry-twisted boundary conditions, especially for the Roberge-Weiss phase transition and the $\mathbb{Z}_N$-QCD. 
These theories has the center-related discrete symmetry, and it is spontaneously broken at high-temperature phases. 
We find that the domain wall connecting distinct states related by the broken center symmetry supports $U(N-1)$ gauge theory with $2$d massless Dirac fermions by explicit weak-coupling computation at sufficiently high temperatures. 

These domain-wall localized theories has the chiral flavor symmetry with an 't~Hooft anomaly. Since 't~Hooft anomaly is a topological object, we argue the persistence of gappless excitations on the domain wall even in the strongly-coupled region of the QCD phase diagram as long as the center symmetry is spontaneously broken. 
We prove this statement using the recent developments about the relation between 't~Hooft anomaly and SPT orders with anomaly-inflow mechanism. 
In other words, we give an interpretation of the pure states related by the broken center symmetry as different SPT orders protected by chiral symmetry. 

\acknowledgments

The work of H.~N. and Y.~T. were supported by Special Postdoctoral Researchers Program of RIKEN. After April, the work of Y.~T. was supported by JSPS Overseas Fellowships. 
\appendix

\section{Domain-wall Dirac fermions and chirality}
\label{sec:domain_wall_chirality}

First, we clarify our convention of the gamma matrices in four dimensions. We only consider the flat Euclidean spacetime. 

In four dimensions, the gamma matrices are the $4\times 4$ matrices $\gamma_{\mu}$ ($\mu=1,\ldots,4$), satisfying 
\be
\{\gamma_{\mu},\gamma_{\nu}\}=2\delta_{\mu\nu}\bm{1}_4. 
\ee
The $\gamma_5$ matrix is introduced by $\gamma_5=\gamma_1\gamma_2\gamma_3\gamma_4$. 
We realize this algebra by the chiral representation,
\be
\gamma_I=\begin{pmatrix}
0&\sigma_I\\
\sigma_I&0
\end{pmatrix},\; 
\gamma_4=\begin{pmatrix}
0&\im \bm{1}_2\\
-\im\bm{1}_2 & 0
\end{pmatrix}, 
\ee
where $\sigma_I$ ($I=1,2,3$) are the $2\times 2$ Pauli matrices. The $\gamma_5$ matrix is expressed by the diagonal matrix in this representation,
\be
\gamma_5=\gamma_1\gamma_2\gamma_3\gamma_4=\begin{pmatrix}
\bm{1}_2&0\\
0&-\bm{1}_2
\end{pmatrix}. 
\ee
The left- and right-handed spinors are defined by the projectors $P_\rmL=(1-\gamma_5)/2$ and $P_\rmR=(1+\gamma_5)/2$, respectively. 
Therefore, the four-component Dirac fermion $\psi$ is represented as $\psi=(\psi_\rmR,\psi_\rmL)=(\psi_{\rmR+},\psi_{\rmR-},\psi_{\rmL+},\psi_{\rmL-})$. 
That is, the first two components are right-handed and the last ones are left-handed. 

We now consider the domain wall fermion. Our set up is that the fourth direction is compactified $\tau\sim \tau+\beta$, and $\beta$ is sufficiently small. The domain wall is set at $x_3=0$ along the $x_1$-$x_2$ directions. 
Using the real mass function $m(x_3)$, the domain wall fermion is obtained as the zero-mode solution of the Dirac equation, 
\be
[\gamma_3 \p_3+\gamma_4 \im m(x_3)]\psi=0. 
\ee
We can easily solve this equation of motion as 
\be
\psi(x_3)=\exp\left[-\im \gamma_3\gamma_4\int_0^{x_3}m(s)\diff s\right]\psi(0). 
\ee
Here, $\im\gamma_3\gamma_4=\mathrm{diag}(-\sigma_3,\sigma_3)=\mathrm{diag}(-1,1,1,-1)$. 

When $m(+\infty)>0$ and $m(-\infty)<0$, the normalizability requires that the first and fourth components must vanish, so that the normalizable zero-mode is given by $\psi=(0,\psi_{\rmR-},\psi_{\rmL+},0)$. 
This is the two-dimensional Dirac fermion on the domain wall, and the chirality of two-dimensions and that of four-dimensions are flipped. 

When $m(+\infty)<0$ and $m(-\infty)>0$, the normalizability requires that the second and third components must vanish, so that the normalizable zero-mode is given by $\psi=(\psi_{\rmR+},0,0,\psi_{\rmL-})$. 
This is also the two-dimensional Dirac fermion on the domain wall, but the chirality of two-dimensions and that of four-dimensions are the same.

\section{Justification of the ansatz of the domain wall}
\label{sec:ansatz_domain_wall}

In this appendix, we will show that the ansatz for the domain wall is correct. This is partly discussed in Ref.~\cite{Bhattacharya:1992qb} for $N=3$ and $N=\infty$, and we here provide the discussion for general values of $N$. 
For simplicity, we restrict our attention to the pure gluon potential in this Appendix. 

We take the following basis of the Cartan matrices of the $\mathfrak{su}(N)$ Lie algebra: 
\bea
&&H_1=\mathrm{diag}(1,-1,0,\ldots,0),\; H_2=\mathrm{diag}(0,1,-1,\ldots,0),\ldots,\; H_{N-2}=\mathrm{diag}(0,\ldots,1,-1,0),\nonumber\\
&&T_{N-1}={2\pi\over N}\mathrm{diag}(1,\ldots, 1, -(N-1)). 
\eea
This satisfies 
\bea
\tr(H_i H_j)&=&2\delta_{i\; j}-\delta_{i+1\; j}-\delta_{i\; j+1}, \nonumber\\
\tr(H_i T_{N-1})&=&0.
\eea
The matrix $\tr(H_i H_j)$ is the tridiagonal Toeplitz matrix.
It is a positive matrix, and its eigenvalues are given as $2+2\cos\left({\pi k\over N-1}\right)>0$ with $k=1,\ldots, N-2$. Another important property in this Cartan basis is that
\be
H_i T_{N-1}={2\pi\over N} H_i. 
\ee
In this basis, we can write the Polyakov loop as 
\bea
\Phi&=& \exp\left(\im \vec{\theta}\cdot \vec{H}+\im \rho T_{N-1}\right)\nonumber\\
&=&\mathrm{diag}\left(\rme^{\im (\theta_1+(2\pi/N)\rho)},\rme^{\im(\theta_2-\theta_1+(2\pi/N)\rho)},\ldots, \rme^{\im(-\theta_{N-2}+(2\pi/N)\rho)}, \rme^{-2\pi\im(N-1)\rho/N}\right),  
\eea
where $\vec{\theta}=(\theta_1,\ldots, \theta_{N-2})$ and $\vec{\theta}\cdot \vec{H}=\theta_1 H_1 +\cdots +\theta_{N-2}H_{N-2}$. 

We note that the gradient of $\tr(\Phi^n)$ vanishes at $\vec{\theta}=0$: 
\be
\left.{\p\over \p \theta_i}\tr(\Phi^n)\right|_{\vec{\theta}=0}=\im n\, \tr\Bigl(H_i \exp(\im n \rho T_{N-1})\Bigr)=0. 
\ee
Let us also compute the Hesse matrix, then we get  
\be
\left.{\p^2\over \p \theta_i \p \theta_j}\tr(\Phi^n)\right|_{\vec{\theta}=0}=-n^2 \tr\Bigl(H_i H_j \exp(\im n \rho T_{N-1})\Bigr)=-n^2 \rme^{2\pi \im n \rho/N}\tr(H_i H_j). 
\ee
Using these properties, we show that at each fixed $\rho$ a local minimum of the $1$-loop Polyakov-loop potential (\ref{V_1loop}) locates at $\vec{\theta}=0$. 
First we take
a derivative of $V$ with respect to $\theta_i$,
\begin{equation}
\frac{\partial}{ \partial \theta_i} V
=
-\frac{2i}{\pi^2 \beta^4} \sum_{n \neq 0} \frac{\mathrm{tr_c} \left(H_i \Phi^n \right)}{n^3} \mathrm{tr_c} \Phi^{-n},
\end{equation}
and find that it is zero at $\vec{\theta} =0$.
Second we compute the Hesse matrix,
\begin{equation}
 \frac{\partial^2}{ \partial \theta_i \partial \theta_j } V
=
\frac{2}{\pi^2 \beta^4} \sum_{n \neq 0} \frac{\mathrm{tr_c} \left(H_i H_j  \Phi^n \right)}{n^2} \mathrm{tr_c} \Phi^{-n} 
-\frac{2}{\pi^2 \beta^4} \sum_{n \neq 0} \frac{\mathrm{tr_c} \left(H_i \Phi^n  \right) \mathrm{tr_c} \left(H_j \Phi^{-n}  \right) }{n^2} .
\end{equation}
At $\vec{\theta} =0$, it becomes
\begin{equation}
\left. \frac{\partial^2 }{ \partial \theta_i \partial \theta_j } V \right|_{\vec{\theta} =0} 
=
\left[
\frac{4}{\pi^2 \beta^4} \sum^{\infty}_{n =1} \frac{N-1+ \cos \left( 2 \pi n \rho  \right) }{n^2}  
\right]
\tr \left( H_i H_j \right) .
\end{equation}
This is a Toeplitz matrix, whose overall factor in the bracket is nonzero and positive for any value of $\rho$.  Therefore $\vec{\theta} = 0$  is a local minimum of the 1-loop Polyakov loop potential for any $N$.

\bibliographystyle{utphys}
\bibliography{./QFT,./ref}
\end{document}